\documentclass{PoS}

\title{Probing New Physics in Charm Couplings with Kaon and Other Hadron Processes}

\ShortTitle{Probing New Physics in Charm Couplings}

\author{\speaker{Jusak Tandean}{\tiny\,}\thanks{This research was supported in part by NSC and NCTS.}\\
        Department of Physics, National Taiwan University, \\ Taipei 106, Taiwan \\
        E-mail: \email{jtandean@phys.ntu.edu.tw}}

\abstract{It is possible that the low-energy effects of physics beyond the standard model can
be parametrized mainly by anomalous couplings of quarks to the $W$ boson.
Such couplings can generate potentially significant contributions to various transitions that
can be probed by current and future experiments.
This work explores constraints on anomalous charm-$W$ couplings from a number of
$CP$-conserving and -violating processes involving the kaon and other flavored hadrons.}

\FullConference{2009 KAON International Conference\\
        June 09 - 12, 2009 \\
        Tsukuba, Japan}

\begin{document}


Current data on low-energy transitions involving the kaon and other flavored hadrons
have confirmed the loop-induced flavor-changing neutral current (FCNC) picture of the standard
model~(SM) and the unitarity of the Cabibbo-Kobayashi-Maskawa (CKM) matrix with three
generations~\cite{pdg}.
However, there is a growing realization that the present understanding of the dynamics of
flavor is incomplete and that physics beyond the SM may be detected in the near future.
The continuing study of FCNC processes with increased precision will, therefore, play
a crucial role in the search for new physics.

In many scenarios of new physics, the new particles are heavier than their SM counterparts and
their effects can be described by an effective low-energy theory.
It is possible that the main effect of such new physics is to modify the SM couplings
between gauge bosons and certain fermions~\cite{Peccei:1989kr}.
The case of anomalous top-quark couplings has been treated before  in
the literature~\cite{Fujikawa:1993zu}, and it was found that they are stringently constrained by
the \,$b\to s\gamma$\, decay.  Interestingly, this mode does not place severe restrictions on
anomalous charm-quark couplings because of the relative smallness of the charm mass.
Here we present the results of a recent study on the possibility of new physics affecting primarily
the charged weak currents involving the charm quark~\cite{He:2009hz}.

The effective Lagrangian in the unitary gauge for a~general parametrization
of anomalous interactions of the $W$ boson with an up-type quark $U_k^{}$ and a~down-type quark
$D_l^{}$ can be written as
\begin{eqnarray} \label{ludw}
{\cal L}_{UDW}^{} \,\,=\,\,  -\frac{g}{\sqrt2}\, V_{kl}^{}\,\bar U_k^{}\gamma^\mu \bigl[
\bigl(1+\kappa_{kl}^{\rm L}\bigr)P_{\rm L}^{}
\,+\, \kappa_{kl}^{\rm R}\,P_{\rm R}^{} \bigr] D_l^{}\, W_\mu^+
\,\,+\,\, {\rm H.c.} \,\,,
\end{eqnarray}
where $g$ is the weak coupling constant, the anomalous couplings $\kappa^{\rm L,R}_{kl}$
are normalized relative to the usual CKM-matrix elements $V_{kl}^{}$, and
\,$P_{\rm L,R}^{}=\frac{1}{2}(1\mp\gamma_5^{})$.\,
Thus the SM limit corresponds to  \,$\kappa^{\rm L,R}_{kl}\to0$.\,
In general, $\kappa^{\rm L,R}_{kl}$ are complex and, as such, provide new sources of $CP$
violation.

\begin{figure}[b]
\includegraphics[width=5in]{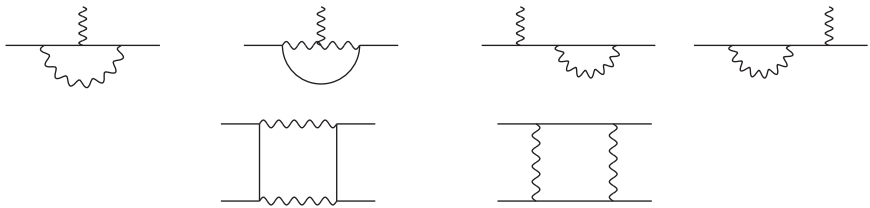}
\caption{\label{loops}
Diagrams contributing to the loop-induced processes discussed in the text.
Straight lines (external wavy lines) denote fermions (neutral gauge bosons),
and the loops contain $W$ bosons besides fermions.}
\end{figure}

These new couplings contribute to flavor-changing processes at one-loop level
and, therefore, affect loop-generated transitions, such as \,$K\to\pi\nu\bar\nu$,\,
\,$K_L\to\ell^+\ell^-$,\, and neutral-meson mixing.
The relevant loop diagrams are displayed in Fig.~\ref{loops}.\footnote{The details of
our loop calculation are presented in Ref.~\cite{He:2009rz}.}
Such processes can be used to place bounds on anomalous couplings in the charm sector.
Below we focus on kaon transitions, but will briefly discuss loop-induced, as well as tree-level,
transitions involving other flavored hadrons.

Since the effective theory with anomalous couplings is not renormalizable, there are
divergent contributions to some of the transitions we consider. Regarding these divergences
as contributions to the coefficients of higher-dimension operators in the theory, we handle
them by following the common procedure of using dimensional regularization, dropping
the resulting pole in four dimensions, and identifying the renormalization scale~$\mu$
with the scale $\Lambda$ of the new physics underlying the effective theory.
Our results thus contain a logarithmic term of the form  $\ln\bigl(\mu/m_W^{}\bigr)$ in
which we set \,$\mu=\Lambda=1$\,TeV\, for definiteness.
In addition, we keep in our estimates those finite terms that correspond to contributions
from SM quarks in the loops.
By adopting this procedure, we trade the possibility of obtaining precise predictions in
specific models for order-of-magnitude estimates of the effects of new physics parametrized
in a model-independent way.

The first transition we consider is the decay \,$K^+\to\pi^+\nu\bar\nu$.\,
To quantify the contribution of the anomalous charm couplings to this mode, it is convenient to
compare it with the dominant contribution in the SM, which comes from the top loop, and
to neglect the masses of the leptons associated with the neutrinos in the new contribution,
so that we can work with just one of them.
The total amplitude can be written in terms of the dominant SM amplitude as~\cite{He:2009hz}
\begin{eqnarray}
{\cal M}(K^+\to\pi^+\nu\bar\nu) \,\,=\,\,
(1+\delta)\,{\cal M}_{\rm SM}^{}\bigl(K^+\to\pi^+\nu\bar\nu\bigr) \,\,,
\end{eqnarray}
\begin{eqnarray}
\delta \,\,=\,\, \frac{V_{cd}^{}V_{cs}^*}{V_{td}^{}V_{ts}^*}\,
\frac{\bigl(\kappa_{cd}^{\rm L}+\kappa_{cs}^{\rm L*}\bigr)
\bigl[-3\,\ln\bigl(\Lambda/m_W^{}\bigr)+4X_0^{}\bigl(x_c^{}\bigr)\bigr]}{4 X\bigl(x_t^{}\bigr)} \,\,,
\end{eqnarray}
where we have kept terms to linear order in the $\kappa$'s, \,$x_q^{}=m_q^2/m_W^2$,\,
$X_0^{}\bigl(x_t^{}\bigr)$ is a loop function, and \,$X\bigl(x_t^{}\bigr)\simeq1.4$\, the QCD-corrected
value of $X_0^{}$~\cite{Buras:2006gb}.
The SM prediction for the branching ratio is
\,${\cal B}_{\rm SM}^{}(K^+\to\pi^+\nu\bar\nu)=(8.5\pm0.7)\times10^{-11}$\,~\cite{Buras:2006gb},
to be compared with its experimental value
\,${\cal B}_{\rm exp}^{}=\bigl(1.73_{-1.05}^{+1.15}\bigr)\times10^{-10}$\,~\cite{Artamonov:2008qb}.
Accordingly, we require  \,$-0.2\le{\rm Re}\,\delta\le 1$,\,  which leads to~\cite{He:2009hz}
\begin{eqnarray} \label{k2pnn}
-2.5\times10^{-4} \,\,\le\,\, -{\rm Re}\bigl(\kappa^{\rm L}_{cd}+\kappa^{\rm L}_{cs}\bigr) \,+\,
0.42\,{\rm Im}\bigl(\kappa^{\rm L}_{cd}-\kappa^{\rm L}_{cs}\bigr) \,\,\le\,\, 1.3\times10^{-3} \,\,.
\end{eqnarray}

The next decay of interest is \,$K_L\to\mu^+\mu^-$.\,
The dominant part of the short-distance contribution to the SM amplitude for this mode
is again induced by the top loop.
Including the anomalous charm contribution at linear order in $\kappa$,
the total short-distance amplitude is~\cite{He:2009hz}
\begin{eqnarray}
{\cal M}_{\rm SD}^{}\bigl(K_L^{}\to\mu^+\mu^-\bigr) \,\,=\,\,
(1+\delta')\, {\cal M}_{\rm SM}^{\rm SD}\bigl(K_L^{}\to\mu^+\mu^-\bigr)  \,\,,
\end{eqnarray}
\begin{eqnarray}
\delta' \,\,=\,\,
\frac{{\rm Re}\bigl[V_{cd}^*V_{cs}^{}\,\bigl(\kappa_{cs}^{\rm L}+\kappa_{cd}^{\rm L*}\bigr)\bigr]
\bigl[-3\,\ln\bigl(\Lambda/m_W^{}\bigr)+4 Y_0^{}\bigl(x_c^{}\bigr)\bigr]}
{4\,{\rm Re}\bigl(V_{td}^*V_{ts}^{}\bigr)\, Y\bigl(x_t^{}\bigr)} \,\,,
\end{eqnarray}
where $Y_0^{}\bigl(x_t^{}\bigr)$ is a loop function and \,$Y\bigl(x_t^{}\bigr)\simeq0.95$\,
its QCD-corrected value~\cite{Gorbahn:2006bm}.
Since the measured branching ratio,
\,${\cal B}\bigl(K_L^{}\to\mu^+\mu^-\bigr)=(6.84\pm0.11)\times10^{-9}$\,~\cite{pdg},  is almost
saturated by the absorptive part of the long-distance contribution,
\,${\cal B}_{\rm abs}^{}=(6.64\pm0.07)\times10^{-9}$\,~\cite{Littenberg:2008zz}, the difference
between them suggests the allowed room for new physics,  \,${\cal B}_{\rm NP}^{}\lesssim3.8\times10^{-10}$,\,
the upper bound being about one half of the SM short-distance contribution,
\,${\cal B}_{\rm SM}^{\rm SD}=(7.9\pm1.2)\times 10^{-10}$\,~\cite{Gorbahn:2006bm}.
Consequently, we demand \,$|\delta'|\le 0.2$,\, which implies~\cite{He:2009hz}
\begin{eqnarray} \label{k2ll}
\Bigl|{\rm Re}\bigl(\kappa_{cs}^{\rm L}+\kappa_{cd}^{\rm L}\bigr) \,+\,
6\times10^{-4}\,{\rm Im}\bigl(\kappa_{cs}^{\rm L}-\kappa_{cd}^{\rm L}\bigr)\Bigr|
\,\,\le\,\, 1.5\times 10^{-4} \,\,.
\end{eqnarray}

We turn now to constraints from $K$-$\bar K$ mixing.
The contribution of the anomalous charm couplings to the matrix element $M_{12}^K$ for the mixing
is given by~\cite{He:2009hz}
\begin{eqnarray} \label{mk}
M_{12}^{K,\kappa} &=& \frac{G_{\rm F}^2\,m_W^2}{24\pi^2}\, f_K^2 m_K^{}\,
V_{cd}^*V_{cs}^{} \left[
\bar\eta^3 B_K^{}\, \bigl(\kappa_{cd}^{\rm L*}+\kappa_{cs}^{\rm L}\bigr)
\left( -V_{td}^*V_{ts}^{}\, x_t^{}\,\ln\frac{\Lambda^2}{m_W^2}
- \sum_q V_{qd}^*V_{qs}^{}\, {\cal B}_1^{}\bigl(x_q^{},x_c^{}\bigr) \right) \right.
\nonumber \\ && \hspace*{14ex} +\, \left.
\frac{\bar\eta^{3/2} B_K^{}\, m_K^2}{\bigl(m_d^{}+m_s^{}\bigr)^2}\,
\kappa_{cd}^{\rm R*}\kappa_{cs}^{\rm R} \left( V_{td}^*V_{ts}^{}\,x_t^{}\,\ln\frac{\Lambda^2}{m_W^2}
+ \sum_q V_{qd}^*V_{qs}^{}\, {\cal B}_2^{}\bigl(x_q^{},x_c^{}\bigr) \right) \right] ~,
\end{eqnarray}
where $\bar\eta$ is a QCD-correction factor and ${\cal B}_{1,2}$ are loop functions, with further
details given in Ref.~\cite{He:2009hz}.
The $K_L^{}$-$K_S^{}$ mass difference  $\Delta M_K^{}$ is related to
\,$M_{12}^K=M_{12}^{K,\rm SM}+M_{12}^{K,\kappa}$\,  by
\,$\Delta M_K^{}=2\,{\rm Re}\,M_{12}^K+\Delta M_K^{\rm LD}$,\,  the long-distance term
$\Delta M_K^{\rm LD}$ being sizable~\cite{Buchalla:1995vs}.
Since the LD part has significant uncertainties, we constrain the anomalous couplings
by requiring that their contribution to $\Delta M_K^{}$ be less than the largest SM SD
contribution, which comes from the charm loop~\cite{Buchalla:1995vs}.
The result is~\cite{He:2009hz}
\begin{eqnarray} \label{kk1}
\bigl|0.043\, {\rm Re}\bigl(\kappa_{cd}^{\rm L}+\kappa_{cs}^{\rm L}\bigr)
+ 0.015\,{\rm Im}\bigl(\kappa_{cd}^{\rm L}-\kappa_{cs}^{\rm L}\bigr)
- {\rm Re}\bigl(\kappa_{cd}^{\rm R*}\kappa_{cs}^{\rm R}\bigr)
+ 0.28\,{\rm Im}\bigl(\kappa_{cd}^{\rm R*}\kappa_{cs}^{\rm R}\bigr) \bigr|
\,\le\, 8.5\times10^{-4} \,\,. \;\;\;
\end{eqnarray}
A complementary constraint on the couplings can be obtained from the $CP$-violation
parameter~$\epsilon$.  It is related to $M_{12}^K$ by
\,$\sqrt2\, |\epsilon|\simeq
\bigl|{\rm Im}\,M_{12}^K\bigr|/\Delta M_K^{\rm exp}$\,~\cite{Buchalla:1995vs},
where \,$\Delta M_K^{\rm exp}=(3.483\pm0.006)\times10^{-15}$\,GeV\,~\cite{pdg} and the small
term containing the $CP$-violating phase in the \,$K\to\pi\pi$\, amplitude has been dropped.
Since  \,$|\epsilon|_{\rm exp}^{}=(2.229\pm0.012)\times10^{-3}$\,~\cite{pdg} and
\,$|\epsilon|_{\rm SM}=\bigl(2.06^{+0.47}_{-0.53}\bigr)\times10^{-3}$\,~\cite{ckmfit},
we require  \,$|\epsilon|_\kappa^{}<0.7\times10^{-3}$\,  for the contribution in Eq.~(\ref{mk}).
This translates into~\cite{He:2009hz}
\begin{eqnarray} \label{kk2}
\bigl|0.015\, {\rm Re}\bigl(\kappa_{cs}^{\rm L}+\kappa_{cd}^{\rm L}\bigr)
+ 0.043\,{\rm Im}\bigl(\kappa_{cs}^{\rm L}-\kappa_{cd}^{\rm L}\bigr)
- 0.28\,{\rm Re}\bigl(\kappa_{cd}^{\rm R*}\kappa_{cs}^{\rm R}\bigr)
- {\rm Im}\bigl(\kappa_{cd}^{\rm R*}\kappa_{cs}^{\rm R}\bigr) \bigr|
\le 2.5\times10^{-6} \,\,. \;\;\;
\end{eqnarray}

The anomalous charm couplings also contribute via gluonic dipole operators to
$\epsilon$ and the $CP$-violation parameter $\epsilon'$ in kaon decay,
as well as to $CP$ violation in hyperon nonleptonic decays~\cite{cmo}.
These operators are generated by the upper diagrams in Fig.~\ref{loops}.
The flavor-conserving counterparts of the gluonic (and electromagnetic) dipole operators
contribute to the electric dipole moment of the neutron~\cite{He:1989xj}.
From the corresponding experimental data, we extract~\cite{He:2009hz}
\begin{eqnarray}
\bigl|{\rm Im}\,\kappa_{cd}^{\rm R}\bigr|\;\lesssim\; 2\times 10^{-3} ~, \hspace{5ex}
\bigl|{\rm Im}\,\kappa_{cs}^{\rm R}\bigr|\;\lesssim\; 2\times 10^{-3} ~.
\end{eqnarray}

Now we briefly discuss constraints from loop-induced transitions involving $B_{d,s}^{}$ mesons.
As in the kaon mixing case, the anomalous charm couplings affect $B_d^{}$ mixing.
The matrix element \,$M_{12}^d=M_{12}^{d,\rm SM}+M_{12}^{d,\kappa}$\, for the mixing is
related to the mass difference  \,$\Delta M_d^{}=2\bigl|M_{12}^d\bigr|$\,  between the heavy
and light mass-eigenstates~\cite{Buchalla:1995vs}.
The measured and SM numbers  \,$\Delta M_d^{\rm exp}=(0.507\pm0.005)\,{\rm ps}^{-1}$\,~\cite{pdg}
and \,$\Delta M_d^{\rm SM}=\bigl(0.563^{+0.068}_{-0.076}\bigr)\,{\rm ps}^{-1}$\,~\cite{ckmfit}
are then related by
\,$\Delta M_d^{\rm exp}=\Delta M_d^{\rm SM}\,\bigl|1+\delta_d^{}\bigr|$\, and
\,$\delta_d^{}=M_{12}^{d,\kappa}/M_{12}^{d,\rm SM}$.\,
Accordingly, we impose  \,$-0.2\le{\rm Re}\,\delta_d^{}\le+0.02$,\,  which leads to~\cite{He:2009hz}
\begin{eqnarray} \label{bb1}
-0.031\,\,\le\,\,\, {\rm Re}\bigl(\kappa_{cb}^{\rm L}+\kappa_{cd}^{\rm L}\bigr) \,+\,
0.4\,{\rm Im}\bigl(\kappa_{cb}^{\rm L}-\kappa_{cd}^{\rm L}\bigr) \,\,\le\,\, 0.003 \,\,.
\end{eqnarray}
An additional constraint can be determined from the measurement of $\beta$ parametrizing
mixing-induced $CP$ violation in \,$B\to J/\psi K$.\,
The anomalous couplings enter via both the mixing and decay amplitudes.
Upon comparing the effective experimental value \,$2\beta^{\rm eff}=0.717\pm0.033$\,~\cite{hfag}
with the SM prediction
\,$2\beta^{\rm SM}=0.753_{-0.028}^{+0.032}$\,~\cite{ckmfit}, we obtain~\cite{He:2009hz}
\begin{eqnarray} \label{2b1}
-1.5\times10^{-3} \,\,\le\,\, 0.4\, {\rm Re}\bigl(\kappa_{cb}^{\rm L}+\kappa_{cd}^{\rm L}\bigr)
- 0.69\, {\rm Im}\,\kappa_{cb}^{\rm L} + {\rm Im}\,\kappa_{cd}^{\rm L}
- 0.31\, {\rm Im}\,\kappa_{cs}^{\rm L} \,\,\le\,\, 0.012 \,\,.
\end{eqnarray}

There are analogous constraints from the $B_s^{}$ sector.
Using the experimental and SM values
\,$\Delta M_s^{\rm exp}=(17.77\pm0.12)\,{\rm ps}^{-1}$\,~\cite{pdg} and
\,$\Delta M_s^{\rm SM}=\bigl(17.6^{+1.7}_{-1.8}\bigr)\,{\rm ps}^{-1}$\,~\cite{ckmfit},
we arrive at~\cite{He:2009hz}
\begin{eqnarray} \label{bb2}
-0.014\,\,\le\,\,{\rm Re}\bigl(\kappa_{cs}^{\rm L}+\kappa_{cb}^{\rm L}\bigr) \,+\,
0.018\,{\rm Im}\bigl(\kappa_{cs}^{\rm L}-\kappa_{cb}^{\rm L}\bigr) \,\,\le\,\, 0.015 \,\,.
\end{eqnarray}
A complementary constraint is provided by the parameter $\beta_s^{}$ in $B_s^{}$ decay,
analogously to $\beta$ in $B_d^{}$ decay.
In this case, the mode of interest is \,$\bar B_s^0\to J/\psi\phi$,\, which proceeds from
the same \,$b\to sc\bar c$\, transition as  \,$\bar B_d^0\to J/\psi\bar K$.\,
The current disagreement between the SM number
\,$2\beta_s^{\rm SM}=0.03614^{+0.00172}_{-0.00162}$\,~\cite{ckmfit} and the measured value
\,$2\beta_s^{\rm eff}=2\beta_{\psi\phi}^{\rm eff}=0.77_{-0.29}^{+0.37}$ or
$2.36_{-0.37}^{+0.29}$\,~\cite{hfag} can be used to extract~\cite{He:2009hz}
\begin{eqnarray} \label{2b2}
-0.09 \,\,\le\,\, 0.026\, {\rm Re}\bigl(\kappa_{cb}^{\rm L}+\kappa_{cs}^{\rm L}\bigr) \,+\,
{\rm Im}\bigl(\kappa_{cb}^{\rm L}-\kappa_{cs}^{\rm L}\bigr) \,\,\le \,\, 7\times10^{-4} \,\,.
\end{eqnarray}

The anomalous charm couplings affect not only loop-induced amplitudes, but also tree-level ones.
The couplings contribute at tree level to \,$D\to\ell\nu$\, and \,$D_s\to\ell\nu$.\,
Existing experimental and theoretical values of the decay constant $f_D$ are consistent with each
other, but those of $f_{D_s}$ disagree at the 2-sigma level~\cite{pdg,Stone:2008gw}.
This information leads us to find~\cite{He:2009hz}
\begin{eqnarray} \label{recd}
\bigl|{\rm Re}\bigl(\kappa_{cd}^{\rm L}-\kappa_{cd}^{\rm R}\bigr)\bigr| \,\,\le\,\, 0.04 ~,
\hspace{5ex}
0 \,\,\le\,\, {\rm Re}\bigl(\kappa_{cs}^{\rm L}-\kappa_{cs}^{\rm R}\bigr) \,\,\le\,\, 0.1 ~.
\end{eqnarray}
Another process affected by the anomalous charm couplings at tree level is the semileptonic decay
\,$b\to c e^-\bar\nu_e^{}$.\,
Consequently, they would pollute the extraction of $V_{cb}^{}$.
From the experimental results on the exclusive modes \,$\bar B\to(D,D^*)e\bar\nu_e^{}$\,
and the inclusive one~\cite{pdg}, we arrive at~\cite{He:2009hz}
\begin{eqnarray} \label{recb}
-0.13 \,\,\le\,\, {\rm Re}\,\kappa^{\rm R}_{cb} \,\,\le\,\, 0 \,\,.
\end{eqnarray}
Lastly, constraints on the couplings can be obtained from comparing the measurements of
the $CP$-violation parameter $\beta$ in $B\to J/\psi K$ and $B\to\eta_c^{}K$.
The SM predicts the same $\sin(2\beta)$ for the two processes, whereas the present data
for its effective values are
\,$\sin\bigl(2\beta_{\psi K}^{\rm eff}\bigr)=0.657 \pm 0.025$\, and
\,$\sin\bigl(2\beta_{\eta_c^{}K}^{\rm eff}\bigr)=0.93 \pm 0.17$\,~\cite{hfag},
which disagree at the 1.5-sigma level.
This could be due to tree-level effects of the anomalous couplings, which would imply~\cite{He:2009hz}
\begin{eqnarray} \label{im(cb+cs)}
-5\times10^{-4} \,\,\le\,\, {\rm Im}\bigl(\kappa_{cb}^{\rm R}+\kappa_{cs}^{\rm R}\bigr)
\,\,\le\,\, 0.04 \,\,.
\end{eqnarray}

\begin{table}[b]
\medskip \centering \small \renewcommand{\arraystretch}{1.5}
\begin{tabular}{|c||c|}
\hline\hline
$0 \le {\rm Re}\,\kappa_{cd}^{\rm L} \le 1.5\times 10^{-4}$ \, &
$\bigl({\rm Im}\,\kappa_{cd}^{\rm L}=0\bigr)$ \\
$0 \le{\rm Re}\,\kappa_{cs}^{\rm L}\le 1.5\times 10^{-4}$ &
$-6\times 10^{-5} \le {\rm Im}\,\kappa_{cs}^{\rm L} \le 6\times 10^{-5}$ \\
\, $-4\times 10^{-3}\le {\rm Re}\,\kappa_{cb}^{\rm L}\le 3\times 10^{-3}$ \, &
$-0.02\le{\rm Im}\,\kappa_{cb}^{\rm L}\le 7\times10^{-4}$  \\
$-0.04 \le {\rm Re}\,\kappa_{cd}^{\rm R}\le 0.04 $ &
\, $-2\times 10^{-3}\le {\rm Im}\,\kappa_{cd}^{\rm R}\le2\times 10^{-3}$ \, \\
$-0.1\le {\rm Re}\,\kappa_{cs}^{\rm R}\le 0$ &
$-5\times10^{-4}\le{\rm Im}\,\kappa_{cs}^{\rm R}\le2\times 10^{-3}$ \, \\
$-0.13\le {\rm Re}\,\kappa_{cb}^{\rm R}\le 0$ &
$-5\times10^{-4}\le{\rm Im}\,\kappa_{cb}^{\rm R}\le 0.04$ \\
\hline\hline
\end{tabular}
\caption{Constraints on each of the anomalous charm couplings.
\label{t:oneatatime}}
\end{table}

In conclusion, we have explored the phenomenological consequences of anomalous $W$-boson
couplings to the charm quark in a comprehensive way.
Kaon processes can be seen to have yielded some of the strongest bounds on the couplings.
In order to gain more insight into the constraints obtained above, we have derived
from them the ranges corresponding to taking only one anomalous coupling at a time to be
nonzero (and only for the cases of a purely real or a purely imaginary coupling).
They are collected in Table~\ref{t:oneatatime}.\footnote{In determining the left-handed
entries in this table, we have taken into account the fact that
only two relative phases among the three left-handed charm-$W$ couplings are
physical~\cite{He:2009hz} and accordingly chosen \,$\phi_{cd}^{\rm L}=0$.}
This table shows that the resulting constraints on the anomalous charm couplings are,
perhaps surprisingly, comparable or tighter than existing constraints
on anomalous $W$-boson couplings to the top quark.
The table also shows that, in general, the left-handed couplings are much more constrained than
the right-handed couplings. Similarly, the imaginary part of the couplings is more tightly
constrained than the corresponding real part. The biggest deviations allowed by current data
appear in the real part of the right-handed couplings, which can be as large as 10\% of
the corresponding SM couplings.
Finally, our study can also serve as a guide as to which future measurements provide the most
sensitive tests for new physics that can be parametrized with anomalous charm-$W$ couplings.

\end{document}